\documentclass[11pt]{article}

\usepackage{amsmath} 
\usepackage[english]{babel}
\usepackage{amsbsy}
\usepackage{amsfonts}
\usepackage{amssymb}
\usepackage{amsthm}
\usepackage[pdftex]{graphicx}
\usepackage{color}
\usepackage{float}
\usepackage{framed}
\usepackage[hyphens]{url}
\usepackage{mathtools}
\usepackage{color, colortbl}
\usepackage{xcolor}
\usepackage{mathrsfs}
\usepackage{geometry}
\usepackage[T1]{fontenc}
\usepackage[utf8]{inputenc}
\usepackage{authblk}
\usepackage{cases}
\usepackage[numbers]{natbib}
\usepackage{makecell}
\usepackage{algorithm}
\usepackage{caption}
\usepackage{longtable}
\usepackage{subcaption}
\usepackage{ltablex}
\usepackage{wrapfig}
\usepackage{lscape}
\usepackage{rotating}
\usepackage{epstopdf}
\usepackage{algpseudocode}
\usepackage{multirow}
\usepackage{bm}

\usepackage{hyperref}
\hypersetup{
	colorlinks=true,
	linkcolor=blue,
	filecolor=blue,
	citecolor = blue,      
	urlcolor=cyan,
}

\title{\textbf{Solving Lane-Emden-Type Eigenvalue Problems with Physics-Informed Neural Networks}}
\author[1]{LO Joel \thanks{oluwaseyejoel@gmail.com; ljoel@uj.ac.za}}
\affil[1]{Department of Mathematics and Applied Mathematics\\
           University of Johannesburg}

\author{C. Harley \thanks{charley@uj.ac.za}} 
 \affil[2]{Department of Electrical and Electronic Engineering Science\\ Faculty of Engineering \& the Built Environment\\ University of Johannesburg\\	South Africa} 
 
 \author[1]{E. Momoniat \thanks{emomoniat@uj.ac.za}}

 \date{}

\begin{document}

\maketitle


\begin{abstract}
The Lane-Emden equation, a nonlinear second-order ordinary differential equation, plays a fundamental role in theoretical physics and astrophysics, particularly in modeling the structure of stellar interiors. Also referred to as the polytropic differential equation, it describes the behavior of self-gravitating polytropic spheres. In this study, we present a novel approach to the solution of the eigenvalue problem which arises when considering the Lane-Emden equation for $ n = 0, 1, 2, 3, 4$ using Physics-Informed Neural Networks (PINNs). The novelty of this work is that, we not only solve the Lane-Emden equation via PINNS but we also determine the eigenvalue, $r$,  which is the stellar radius. Hyperparameter tuning was conducted using Bayesian optimization in the Optuna framework to identify optimal values for the number of hidden layers, number of neurons, activation function, optimizer, and learning rate for each value of $n$. The results show that, for $n = 0, 1$, PINNs achieve near-exact agreement with theoretical eigenvalues (errors < $0.000806\%$). While for more nonlinear cases, $ n = 2, 3$ and $n=4$, PINNs yield errors below $0.0009\%$ and $0.05\%$ respectively, validating their robustness. 
\end{abstract}


\section{Introduction}
The polytropic Lane-Emden equation describes how the pressure and density vary with each other and it is a useful approximation for self-gravitating spheres of plasma such as stars \cite{chandrasekhar1957, horedt2004} . For the derivation of the Lane-Emden Equation, see the following literature \cite{mukherjee2011, hansen2012, el2023b}. The solutions to the Lane-Emden equations help researchers gain insights into star formation, thermodynamics, galaxy clusters, and other fundamental aspects of astrophysics, making it a key tool in the theoretical framework of stellar structure and evolution. This study uses the state-of-art method - physics-informed neural networks (PINNs) - to obtain the solution of the polytropic Lane-Emden Equation and the value of its first zero.

The polytropic Lane-Emden equation \cite{chandrasekhar1957, horedt2004, momoniat2006, harley2008, harley2008b} reads 
\begin{equation}
	\frac{1}{\xi^2} \frac{d}{d \xi} \left( \xi^2 \frac{d \theta}{d \xi}  \right) + \theta^n = 0
	\label{eqn1}
\end{equation}
where $ \xi$ is a dimensionless radius,  $\theta$ is related to the density by the following equation  
\begin{equation}
	\rho =\rho_{c} \theta ^{n}
\end{equation} 
for central density $ \rho _{c}$. The index $n$ is the polytropic index that appears in the polytropic equation of state, 
\begin{equation}
P= K \rho ^{1+{\frac {1}{n}}},
\end{equation} 
where $P$  and $ \rho $ are the pressure and density, respectively, and $ K $ is a constant of proportionality. 

Equation (\ref{eqn1}) can be rewritten as 
\begin{equation}
	 \frac{d^2 \theta}{d \xi^2} + \frac{2}{\xi} \frac{d \theta}{d\xi} + \theta^n = 0
	\label{eqn2}
\end{equation}
which is a second order nonlinear differential equation (DE) with these two boundary conditions: 
\begin{equation}
	   \theta(0) = 1 \quad \text{and} \quad \frac{d \theta}{d \xi}(0) = 0.
	\label{DEcond}
\end{equation} 
The condition $\theta(0) = 1$ represents the fact that at the center, $\xi = 0$, the dimensionless density, $\theta$, of the star (or fluid sphere) is maximum and normalized to 1. And the condition $\frac{d\theta}{d\xi}(0) = 0$ denotes that there is no preferred direction (density gradient is zero) at the center of a spherically symmetric object. 

In this study, we reformulate the Lane-Emden equation in \eqref{eqn2} and \eqref{DEcond} as an eigenvalue problem, and we employ the PINNs methodology when obtaining the solution of the eigenvalue problem which arises when considering the Lane-Emden equation for $ n = 0, 1, 2, 3, 4$. This study does not only solve the Lane-Emden equation via PINNs but it also determine the eigenvalue, $r$,  which is the stellar radius, for each value of $n$. Hyperparameter tuning was conducted using Bayesian optimization in the Optuna framework to identify optimal values for the number of hidden layers, number of neurons, activation function, optimizer, and learning rate for each value of $n$.

The remaining part of this study is organized as follows. In Section~\ref{literature}, we review some related studies that have solved the Lane-Emden equation using different methods. Next, in Section~\ref{eigenvalue}, we show that the Lane-Emden equation could be formulated as an eigenvalue problem. In Section~\ref{pinnss}, we introduce the PINNs approach we are using to solve the eigenvalue problem of the Lane-Emden equation in this study. In Section~\ref{resultss}, the results obtained from our approach are presented. And we conclude the study in Section~\ref{conclude}.

\section{Literature Review}
\label{literature}
The Lane-Emden equation has received much attention from the literature since it seems to be a fundamental tool in astrophysics for modeling the structure and evolution of stars. It has also been used to derive important stellar properties such as mass, radius, and density profiles \cite{chandrasekhar1957}. El-Essawy et al. \cite{el2023} proposed a computational technique based on Monte Carlo algorithms to solve Lane-Emden type equations arising in astrophysics, analyzing four specific equations: positive and negative indices of polytropic gas spheres, isothermal gas sphere, and the white dwarf equation. The Monte Carlo method was compared to numerical and analytical models, showing good agreement for the four Lane-Emden equations studied in the paper. Alves and Rădulescu \cite{alves2020} presented an analysis of the Lane-Emden equation with variable exponent and Dirichlet boundary condition without assuming any subcritical hypotheses which allows the equation to model a wider array of physical or geometric scenarios. The authors included a consideration of mixed regimes of the reaction, covering both radial and non-radial cases, allowing the equation to exhibit behavior that spans across different growth conditions.


Zamiri et al. \cite{zamiri2021} presented the Laguerre collocation method to obtain numerical solutions for both linear and nonlinear Lane-Emden-type equations along with their initial conditions. The method used operational matrices with respect to modified generalized Laguerre polynomials (MGLPs) to transform the main equation and initial conditions into a matrix equation, which allowed for the determination of coefficients of the approximate solution through solving a system of algebraic equations. Kumar et al. \cite{kumar2011} presented a numerical method for solving linear and nonlinear Lane-Emden-type equations using the Bernstein polynomial operational matrix of integration. Some special cases of the Lane-Emden equation were considered to demonstrate the efficiency of the proposed method.

Ahmad et al. \cite{ahmad2016} proposed an hybrid computational methods utilizing unsupervised neural network models and stochastic optimization techniques to solve nonlinear singular Lane-Emden type differential equations arising in astrophysics models. The proposed approximated solutions of higher order ordinary differential equations were calculated using neural networks trained with genetic algorithm and pattern search hybrid with sequential quadratic programming, which showed good agreements with standard solutions.

Mukherjee et al. \cite{mukherjee2011} derived solutions for the Lane-Emden equation for different values of the polytropic index $(n = 0, 1, 2, 3, 4, 5)$ using the differential transform method (DTM), which is an exact series solution method. The authors provided the solutions for each value of $n$, which demonstrated the application of DTM to solve the Lane-Emden nonlinear equation. Kazemi \cite{kazemi2018} presented the numerical solution of the general Lane-Emden equation using a collocation method based on Double Exponential DE transformation. The method was used to convert the equation into a nonlinear Volterra integral equation, and numerical examples demonstrated the accuracy of the method. 

Similar to the work conducted in this research, Baty \cite{baty2023} introduced PINNs to solve the Lane-Emden type equation - polytropic, isothermal and white dwarf cases. The study detailed how PINNs can be used to constrain the equation residuals at specific collocation points, alongside boundary data, through a minimization process. The study also showed the ability of PINNs to learn solutions for multiple equations simultaneously using the same network. This is particularly beneficial for families of equations, such as the polytropic equations for various indices, showcasing the flexibility and efficiency of this method. The results of the PINNs method were compared with two other numerical methods - Monte Carlo and Chebyshev Neural Network methods. The results showed the advantages of PINNs in terms of accuracy and efficiency when solving the Lane-Emden equations. Further work was conducted using this methodology by Mazraeh and Parand \cite{mazraeh2024gepinn} who employed the hybridization of grammatical evolution (GE) algorithm and PINNs to symbolically solve the Lane-Emden equation. The GE algorithm was used to construct mathematical expressions that include various parameters while PINNs was used to determine these parameters. This hybridization allows for a more robust and accurate symbolic solution to nonlinear ordinary differential equations compared to traditional numerical methods.

\section{The Eigenvalue Problem}
\label{eigenvalue}
As noted in the introduction section, Equation \eqref{eqn2} has a constant $n$ that represents the polytropic index. Also, the physical problem that needs to be solved requires us to find a point $r$ such that $\theta(r) = 0$. This is called the first zero of the polytrope function, $\theta$. The value of $r$ is connected to the value of $n$ in a physically meaningful way. That is, for $n \in [0, 4]$, we get different values of eigenvalue, $r$. To show that this is an eigenvalue problem, we make the following coordinate transformation
\begin{equation}
	z = \frac{\xi}{r}
	\label{eqn5}
\end{equation}

with 

\begin{equation}
	\frac{d}{d\xi} = \frac{dz}{d\xi}.\frac{d}{dz} = \frac{1}{r}\frac{d}{dz}, \quad  \frac{d^2}{d\xi^2} = \frac{d}{d\xi} \left( \frac{d}{d\xi} \right) = \frac{1}{r^2}\frac{d^2}{dz^2}
	\label{eqn6}
\end{equation}
when $\xi = 0$, $z = 0$, and when $\xi = r$, $z = 1$. We shall, therefore, attempt to find the solution to 
\begin{align}
	\label{eqn7}
	& \frac{d^2 \theta}{d z^2} + \frac{2}{z} \frac{d \theta}{dz} + r^2 \theta^n = 0, \\ \nonumber
	& \theta(0) = 1, \quad \theta'(0) = 0 \quad  \theta(1) = 0.
\end{align}
It should be noted that $r$ is the eigenvalue. Also, we note that at $z=0$, the equation is singular. As such, we reformulate Equation \eqref{eqn7} so that it can be re-written in the following equivalent form
\begin{align}
	\label{eqn7c}
	& z \frac{d^2 \theta}{d z^2} + 2 \frac{d \theta}{dz} + z r^2 \theta^n = 0, \\ \nonumber
	& \theta(0) = 1, \quad \theta'(0) = 0 \quad  \theta(1) = 0.
\end{align}
In this form, the singularity at $z=0$ which generally poses numerical difficulties in traditional discretization methods disappears \cite{baty2023}. For our method, this form gives a better precision than the form in Equation \eqref{eqn7}. Hence, we take Equation \eqref{eqn7c} as the eigenvalue problem of the Lane-Emden equation, and we shall be using PINNs to determine its solution as well as its eigenvalue, which is the radius of the stellar structure.

Normally, the two boundary conditions in Equation \eqref{DEcond} would be sufficient to specify a unique solution to the problem in Equation \eqref{eqn2}. However,  for the eigenvalue problem in Equation \eqref{eqn7c}, they are not sufficient, since the condition $\theta (1) = 0$ defines an eigenvalue problem for eigenvalue, $r$, ensuring that we can uniquely determine the solution and the parameter. Without it, the problem would be underdetermined in the context of finding the correct eigenvalue, $r$, such that the solution meets the physical. Hence, a third boundary condition is utilized which stipulates that the  dimensionless density is zero at the extreme boundary. Table \ref{exactTable} presents the list of known polytropic indexes, their exact solutions and stellar radius, $r$, in the literature \cite{chandrasekhar1957,baty2023,el2023,mall2014}.
\begin{table}[H]
	\centering
	\caption{The Polytropic Index, Exact Solution and Radius}
	\begin{tabular}{|c|c|c|}
		\hline
		\textbf{Polytropic Index} & \textbf{Exact Solution} & \textbf{Radius} \\
		\hline
		$n=0$ & $\theta(\xi) = 1 - \frac{\xi^2}{6}$ & $r = \sqrt{6}$ \\
		\hline
		$n=1$ & $\theta(\xi) = \frac{\sin \xi}{\xi}$  &  $r = \pi$ \\
		\hline
		$n=5$ & $\theta(\xi) = \left( 1 + \frac{\xi^2}{3} \right)^{-1/2}$ & $r = \infty$ \\
		\hline		
	\end{tabular}
	\label{exactTable}
\end{table}

Most literature \cite{mall2014,he2020deep,singh2018reliable,ahmed2023numerical,kashem2020approximate}, solve the Lane-Emden equation directly, not as an eigenvalue problem. Meaning, they are not solving to obtain the radius of the Lane-Emden equation. However, in \cite{yip2017}, the authors solved a scaled Lane-Emden equation as an eigenvalue problem using a perturbative method. This method involves expanding the solution in a series form around known exact solutions at specific indices (namely $n=0$ and $n=1$). The study observed that the method faced challenges related to the slow convergence of the series solution for certain values of the polytropic index, $n$. Specifically, for $n > 1.9121$ the series solution diverged before reaching the surface of the polytrope, which limited the applicability of the method in these cases \cite{yip2017}. The PINNs method we employ, in this study, does not use a scaled Lane-Emden equation neither does it involve the knowledge of the exact solution. It solves Equation~\eqref{eqn7} to obtain the value of $r$ suitable for each specific polytropic index, $n \in [0, 4]$. The PINNs approach is applicable in all cases. 

\section{Physics Informed Neural Networks}
\label{pinnss}
Physics-informed neural networks (PINNs) \cite{raissi2019, karniadakis2021} are artificial neural networks designed to solve supervised learning tasks while incorporating the governing physical laws described by ordinary or partial differential equations. To make the model aware of these underlying equations, the first step is to define a loss function that captures the constraints imposed by the physics. The PINNs is then trained to directly approximate the solution to the differential equation. From the literature, PINNs has been used to solve both forward and inverse problems \cite{raissi2019}, integro-DE \cite{yuan2022}, fractional DE \cite{pang2019,ren2023} and stochastic DE \cite{yang2020,zhang2020l}.
 
The total loss function for the Lane-Emden equation in \eqref{eqn7} is given as:
\begin{equation}
	L(\alpha) = L_{equation}  + L_{BC1} + L_{BC2} + L_{BC3}
	\label{lossEq}
\end{equation}
where $L_{equation}$, $L_{BC1}$, $L_{BC2}$ and $L_{BC3}$ are the loss for the differential equation, the boundary conditions respectively. The total loss function, $L(\alpha)$, measures how well the neural network $\theta_{PINN}(z;\alpha)$ satisfies the Lane-Emden equation and its boundary conditions, where $\alpha$ denotes the parameters of the network. Each of these terms is as follows.
\begin{equation}
 L_{equation} = \frac{1}{N} \sum_{i=1}^{N} \left[ z \frac{d^2 \theta_{PINN}}{d z^2} + 2 \frac{d \theta_{PINN}}{dz} + z r^2 \theta^n_{PINN}  \right]^2
\end{equation}

\begin{equation}
	L_{BC1} =  \left ( \theta_{PINN} (0) - 1 \right )^2
\end{equation}

\begin{equation}
	L_{BC2} =   \left( \frac{d \theta_{PINN}}{dz} (0) - 0  \right)^2
\end{equation}

\begin{equation}
	L_{BC3} =  \left ( \theta_{PINN} (1) - 0 \right )^2
\end{equation}



\begin{figure}[!htp]
	\centering
	\begin{minipage}{0.85\linewidth}
		\includegraphics[width=\linewidth]{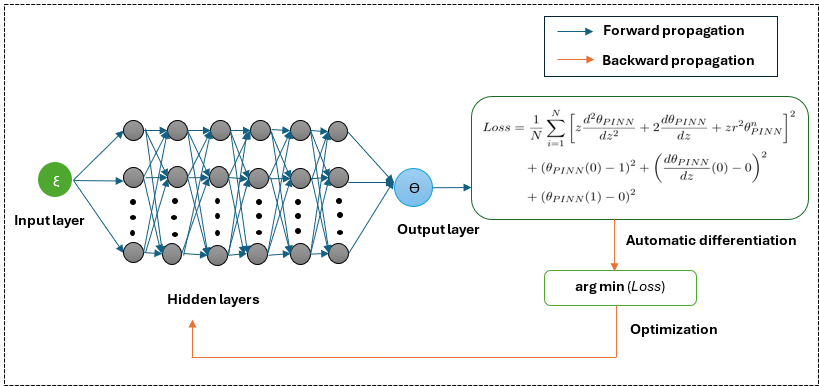}
	\end{minipage}
	\caption{The PINNs Architecture used}
	\label{deepFig}	
\end{figure}

For the PINN solutions of the Lane-Emden equation, we build a fully connected neural
network (see Figure \ref{deepFig}).

\section{Results and Discussion}
\label{resultss}
Using the PINNs approach described in the previous section, we now present, in this section, the eigenvalue solutions of the Lane-Emden equation using PINNs. Hyperparameter tuning was performed, using Bayesian Optimization, for the different values of the number of hidden layers, number of neurons, activation functions, optimizers and the learning rate. In Table~\ref{hyParam}, we provide the different values/options for the hyperparameters and the selected best choices for $n=0, 1, 2, 3, 4$. It should be observed from the table, that Adam optimization was chosen for all the values of $n$, because it produces the lowest loss value when it is employed in the computations. The activation function obtained for $n=0, 1, 2$ is SiLU and for $n=3, 4$, we have the Tanh activation function. The mathematical models of each activation function which were considered during hyperparameter tuning are given below:
\begin{equation}
	\text{Sigmoid}(z) = \frac{1}{1 + e^{-z}},
	\label{eqqn1}
\end{equation}

\begin{equation}
	\text{ReLU}(z) = 
	\begin{cases}
		0 & \text{if } x \leq 0\\
		x & \text{if } x > 0
	\end{cases} \quad = \text{Max}(0, x),
	\label{eqqn1b}
\end{equation}

\begin{equation}
	\text{SiLU}(z) = \frac{z}{1 + e^{-z}},
	\label{eqqn2}
\end{equation}

\begin{equation}
	\text{Tanh}(z) = \frac{e^z - e^{-z}}{e^z + e^{-z}}.
	\label{eqqn3}
\end{equation}
\vspace{0.05cm}

\noindent These are non-linear functions that enable the neural network to learn the complex patterns present in the solution of the Lane-Emden equation. They provide gradients that are important for updating the weights and biases in the neural network learning process, thus enabling backpropagation.

The learning rate for $n=1, 2$ is $0.0001$ while for $n=2, 3, 4$, the learning rate is $0.001$. The loss recorded for the best range of hyperparameters for $n =0, 1, 2, 3, 4$ were $3.7 \times 10^{-5}$, $1.0 \times 10^{-7}$, $1.0 \times 10^{-6}$, $6.6 \times 10^{-5}$ and $ 4.1 \times 10^{-4} $ respectively. The PINNs solution was implemented in Python programming language using pytorch (torch-2.6.0+cu124). Algorithm~\ref{algPinns} describes how we solve the eigenvalue Lane-Emden equation uisng PINNs. Unlike classical numerical methods, PINNs integrate the physical laws (i.e., differential equations and boundary conditions) directly into the loss function of a neural network. Hence, the novelty of this study lies with the ability of PINNs to leverage the underlying physics of the eigenvalue problem of Lane-Emden equation to learn the solutions as well as obtain the eigenvalues.

\begin{table}[thb]
	\centering
	\caption{Hyperparameters determined for $n = 0, 1, 2, 3$ and $4$}
	\begin{tabular}{|c|c|c|c|c|c|c|}
		\hline
		\textbf{Hyperparameters} & \textbf{\makecell{Range of \\ Values/Options}} & $\bm{n=0}$  & $\bm{n=1}$  & $\bm{n=2}$  & $\bm{n=3}$ & $\bm{n=4}$ \\
		\hline
		\textit{\makecell{Number \\ of \\ hidden layers}} & 2, 3, 4, 5 &  3  &  5  & 3 & 5 & 3\\
		\hline
		\textit{\makecell{Number \\ of \\ neurons \\ per layer}} &  20, 30, 40, 50 & 50  & 50 & 30 & 50 & 30 \\
		\hline
		\textit{\makecell{Activation \\ Function} }& \makecell{Tanh, \\ ReLU, \\ Sigmoid, \\ SiLU} & SiLU  &  SiLU & SiLU & Tanh & Tanh \\
		\hline
		\textit{Optimizer} & \makecell{Adam, \\  SGD, \\ RMSprop, \\ Adagrad} &  Adam &  Adam & Adam & Adam & Adam \\
		\hline
		\textit{Learning rate} & \makecell{0.01, \\ 0.05, \\ 0.001, \\ 0.005, \\ 0.0001} & 0.0001 &  0.0001 & 0.001 & 0.001  & 0.001 \\
		\hline	
	\end{tabular}
	\label{hyParam}
\end{table}

The function $\theta(z)$, in Algorithm~\ref{algPinns}, representing the solution of the Lane-Emden equation, is approximated by a neural network  $\theta(z; \mathbf{w})$, where  $\mathbf{w}$ denotes the trainable weights. The eigenvalue, $r$ is treated as a trainable scalar parameter, reparameterized as $r = \exp(r_{\text{raw}})$ to enforce positivity. During training, the neural network model minimizes a composite loss: (1) physics-informed loss which quantifies the deviation from the Lane-Emden differential equation at multiple collocation points in the domain $z \in [0, 1] $, and (2) boundary condition loss which enforces the required physical constraints at the domain boundaries. The algorithm uses automatic differentiation to compute derivatives of the neural network outputs, ensuring precision in evaluating its residuals. Additionally, learning rate scheduling and model check-pointing are used to improve training efficiency and reliability. This approach allows the simultaneous approximation of both the eigenfunction, $\theta(z)$ and the eigenvalue, $r$. 


\begin{algorithm}[htb]
\caption{Solving the Eigenvalue Lane-Emden Equation using PINNs}
\label{algPinns}
\begin{algorithmic}
\State \textbf{Input:} 
\State Power index $ n $, number of collocation points $ N $, learning rate $ \eta $, \\ number of training epochs $ E_{\max} $

\State \textbf{Initialization:} 
\State Initialize neural network $ \theta(z; \mathbf{w}) $ with parameters $ \mathbf{w} $
\State Initialize trainable parameter $ r_{\text{raw}} \in \mathbb{R} $; define $ r = \exp(r_{\text{raw}}) $
\State Sample $ N $ collocation points $ \{ z_i \}_{i=1}^N \in [0, 1] $

\State \textbf{Training Loop:} 
\For{$\text{epoch} = 1$ to $E_{\max}$}
\State \textbf{Compute PDE residual:}
\For{each collocation point $ z_i $}
\State Compute $ \theta(z_i) $, $ \theta'(z_i) $, $ \theta''(z_i) $ using automatic differentiation
\State Compute residual:
\[
R(z_i) =  z_i \theta''(z_i) + 2 \theta'(z_i) +  z_i r^2 \theta(z_i)^n
\]
\EndFor
\State Compute physics-informed loss:
\[
\mathcal{L}_{\text{physics}} = \frac{1}{N} \sum_{i=1}^N \left( R(z_i) \right)^2
\]

\State \textbf{Compute boundary condition loss:}
\[
\mathcal{L}_{\text{bc}} = (\theta(0) - 1)^2 + (\theta'(0)-0)^2 + (\theta(1) - 0)^2
\]

\State \textbf{Total loss:} $ \mathcal{L} = \mathcal{L}_{\text{physics}} + \mathcal{L}_{\text{bc}} $

\State \textbf{Backpropagation:} 
\State Backpropagate gradients of $ \mathcal{L} $ with respect to $ \mathbf{w} $ and $ r_{\text{raw}} $
\State Update parameters using Optimizer
\State \textbf{Learning rate scheduler:} 
\State Adjust learning rate using ReduceLROnPlateau scheduler
\State \textbf{Save the best model:} 
\State Save current best model if $ \mathcal{L} $ is minimized
\EndFor

\State \textbf{Output:} 
\State Trained network $ \theta(z) $, estimated eigenvalue $ r $ and the loss plot
\end{algorithmic}
\end{algorithm}

\begin{figure} 
	\centering
	\begin{subfigure}{.5\textwidth}
		\centering
		\includegraphics[width=.8\linewidth]{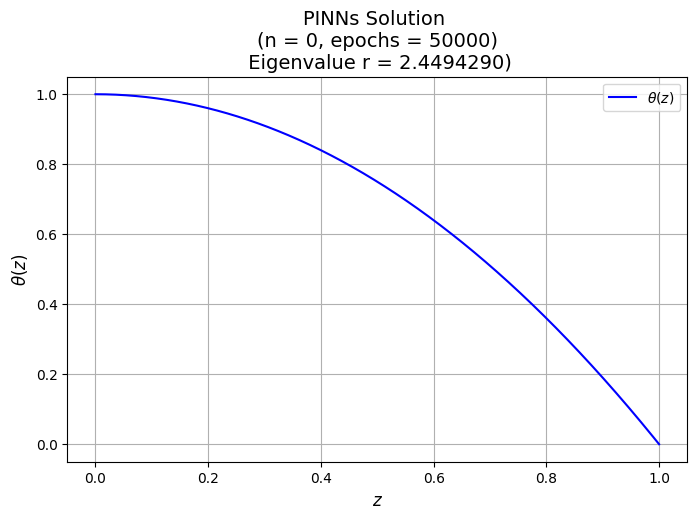}
		\caption{}
		\label{n01a}
	\end{subfigure}%
	\begin{subfigure}{.5\textwidth}
		\centering
		\includegraphics[width=.8\linewidth]{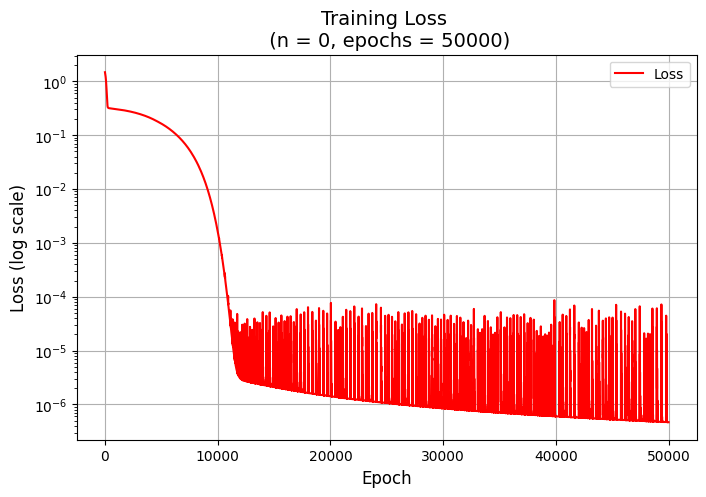}
		\caption{}
		\label{n01b}
	\end{subfigure}
	\caption{Solutions for $n=0$ with 50 000 epochs: (a) The PINNs solution (b) The Training Loss}
	\label{fign0-1}
\end{figure}

Starting from $n=0$, Figure~\ref{fign0-1} shows the results of the PINNs solutions and the training loss plots for it. The PINNs' approximation of the Lane-Emden solution, see Figure \ref{n01a}, shows a smooth decrease from $\theta(0) = 0$ (center of the star) to $\theta(1) = 0$ (surface). The value of $r$ obtained shows the first zero of $\theta(z)$ for $n=0$, which is theoretically given as $\sqrt{6}$, see Table~\ref{exactTable}. The eigenvalues, $r$, obtained compared to the literature is given in Table \ref{eigenV}. Figures~\ref{n01b} shows the training loss on a logarithmic scale for the number of epochs. The loss decreases monotonically, indicating successful training and suggesting that the PINNs effectively learned the underlying physics and the boundary conditions of the problem. The training loss obtained is  $\approx 10^{-8}$.

\begin{figure} 
	\centering
	\begin{subfigure}{.5\textwidth}
		\centering
		\includegraphics[width=.8\linewidth]{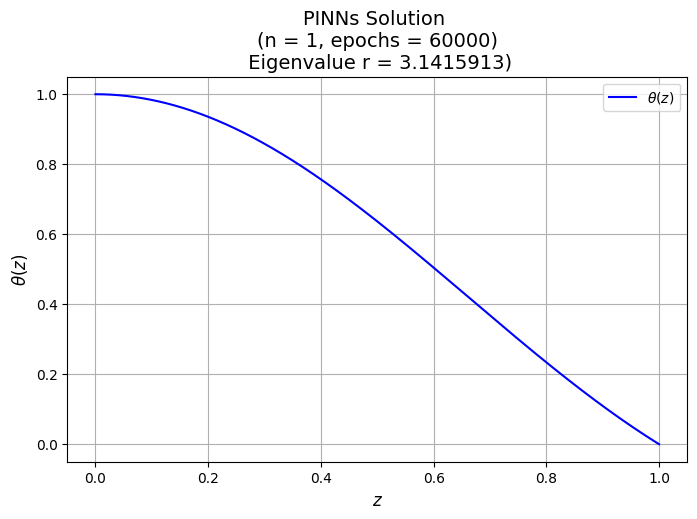}
		\caption{}
		\label{n11a}
	\end{subfigure}%
	\begin{subfigure}{.5\textwidth}
		\centering
		\includegraphics[width=.8\linewidth]{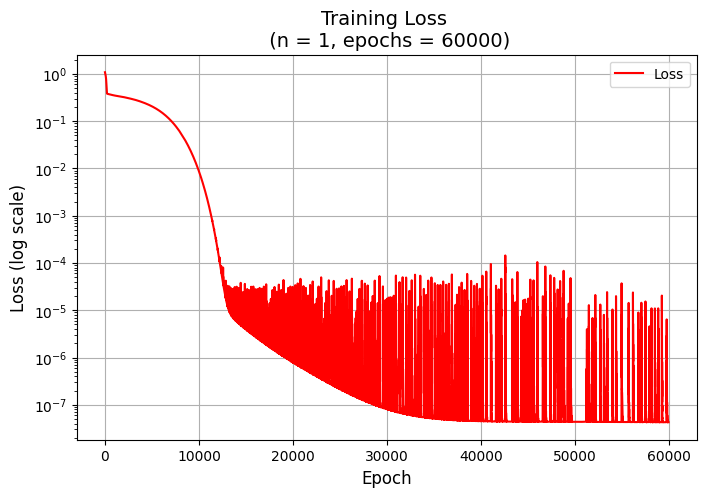}
		\caption{}
		\label{n11b}
	\end{subfigure}
	\caption{Solutions for $n=1$ with 60 000 epochs: (a) The PINNs solution (b) The Training Loss}
	\label{fign1-1}
\end{figure}

\begin{figure} 
	\centering
	\begin{subfigure}{.5\textwidth}
		\centering
		\includegraphics[width=.8\linewidth]{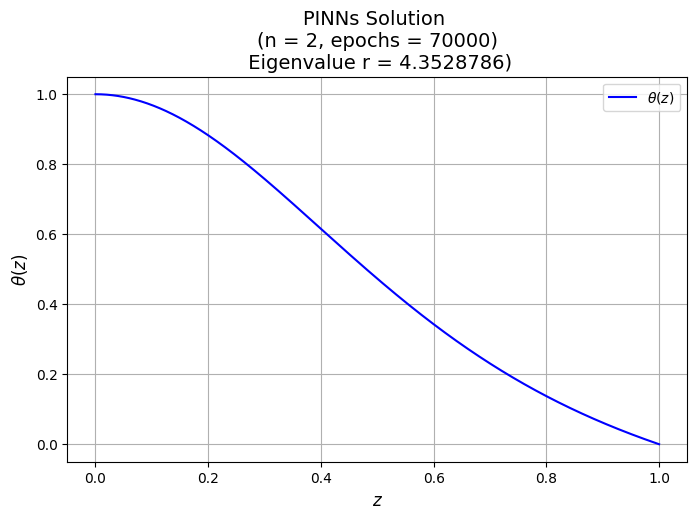}
		\caption{}
		\label{n21a}
	\end{subfigure}%
	\begin{subfigure}{.5\textwidth}
		\centering
		\includegraphics[width=.8\linewidth]{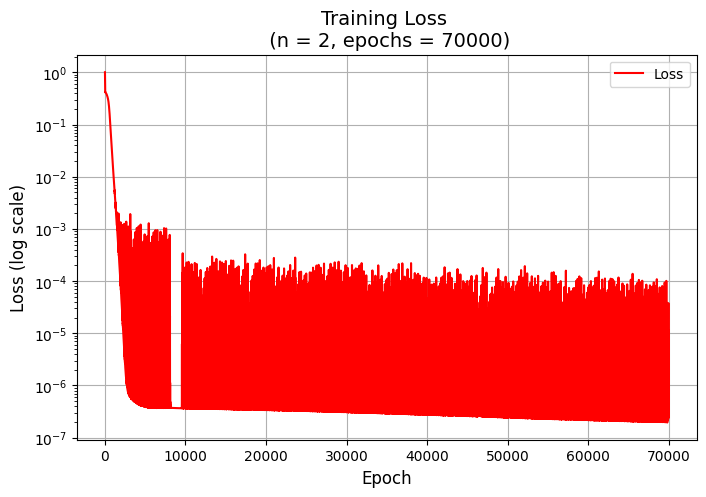}
		\caption{}
		\label{n21b}
	\end{subfigure}
	\caption{Solutions for $n=2$ with 70 000 epochs: (a) The PINNs solution (b) The Training Loss}
	\label{fign2-1}
\end{figure}

\begin{figure} 
	\centering
	\begin{subfigure}{.5\textwidth}
		\centering
		\includegraphics[width=.8\linewidth]{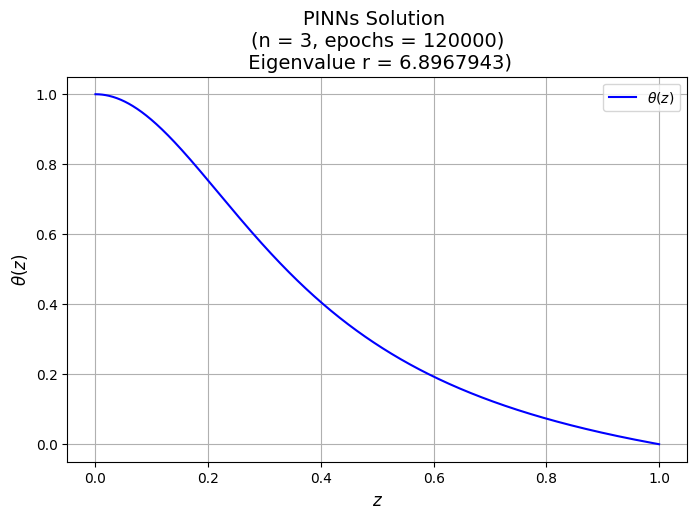}
		\caption{}
		\label{n31a}
	\end{subfigure}%
	\begin{subfigure}{.5\textwidth}
		\centering
		\includegraphics[width=.8\linewidth]{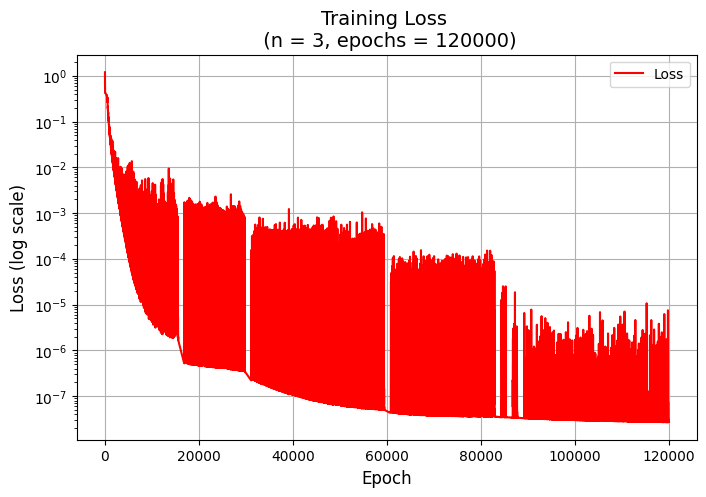}
		\caption{}
		\label{n31b}
	\end{subfigure}
	\caption{Solutions for $n=3$ with 120 000 epochs: (a) The PINNs solution (b) The Training Loss}
	\label{fign3-1}
\end{figure}

\begin{figure} 
	\centering
	\begin{subfigure}{.5\textwidth}
		\centering
		\includegraphics[width=.8\linewidth]{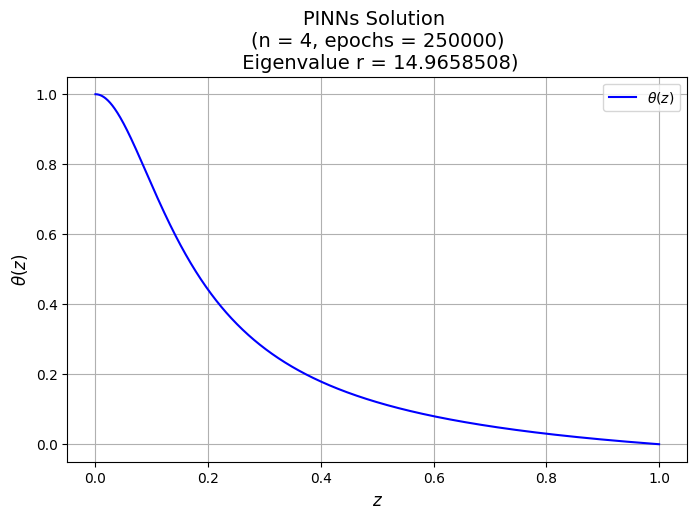}
		\caption{}
		\label{n41a}
	\end{subfigure}%
	\begin{subfigure}{.5\textwidth}
		\centering
		\includegraphics[width=.8\linewidth]{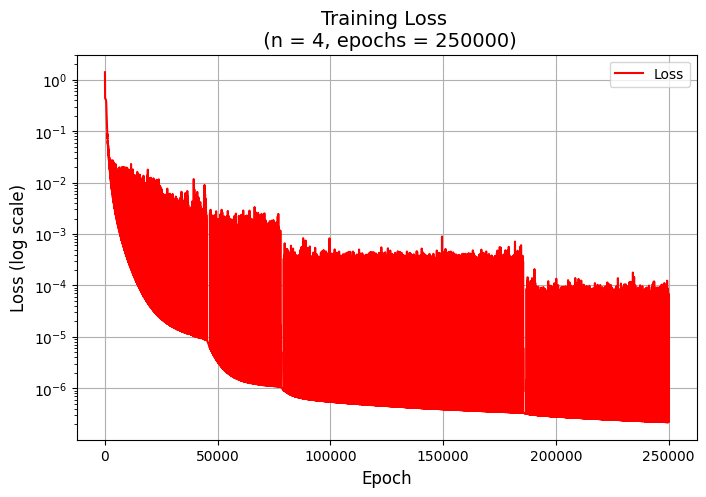}
		\caption{}
		\label{n41b}
	\end{subfigure}
	\caption{Solutions for $n=4$ with 250 000 epochs: (a) The PINNs solution (b) The Training Loss}
	\label{fign4-1}
\end{figure}

In Figure \ref{fign1-1}, we show the results when $n=1$. The PINNs' solutions of the Lane-Emden equation for this case are shown in Figure~\ref{n11a} with the eigenvalues obtained. In Table~\ref{eigenV}, we compare the PINNs' eigenvalue with the one reported in \cite{chandrasekhar1957}. The eigenvalue obtained with PINNs is correct up to $5$ decimal places. The reported value in the literature was only given up to $5$ decimal places. Figure~\ref{n11b} shows the training loss obtained, which is $\approx 10^{-8}$ for $60 000$ epochs. Similarly, the PINNs' solutions for $n=2$, is shown in Figure~\ref{n21a}. While the training losses is shown in Figure~\ref{n21b}. The PINNs results give the correct value up to $5$ decimal places, see Table~\ref{eigenV}. The training loss is  $\approx 10^{-7}$. Also, Figure~\ref{n31a} shows the result for $n=3$ and its associated training loss in Figure \ref{n31b}. Here also, the PINNs' approximation is correct up to $3$ decimal place, see Table~\ref{eigenV}. On the training loss, the values obtained is  $\approx 10^{-8}$. Lastly, for $n=4$, Figure~\ref{n41a} and Figure~\ref{n41b} show the solution and the training loss. The PINNs' approximation, for this polytropic index, is correct up to $1$ decimal place, see Table~\ref{eigenV}. The training loss, as shown in Figure~\ref{n41b}, is $\approx 10^{-7}$.

\begin{table}[H]
	\centering
	\caption{Comparing the PINNs Eigenvalues for each Polytropic Index}
	\begin{tabular}{|c|c|c| c|}
		\hline
		Polytropic Index & PINNs Eigenvalue & Radius \cite{chandrasekhar1957} & Exact \\
		\hline
		$n=0$ & 2.4494290 &  2.4494 & $\sqrt{6} $ = 2.44948974278 \\
		\hline
		$n=1$ & 3.1415913  &  3.14159 & $\pi $  = 3.14159265359 \\
		\hline
		$n=2$ & 4.3528786  & 4.35287 & - \\
		\hline	
		$n=3$ & 6.8967943  & 6.89685 & - \\
		\hline	
		$n=4$ & 14.9658508  & 14.97155 & -  \\
		\hline
	\end{tabular}
	\label{eigenV}
\end{table}

\section{Conclusion}
\label{conclude}
This study demonstrates the application of PINNs in solving the Lane-Emden eigenvalue problem for polytropic indices $ n = 0, 1, 2, 3, 4$. The novelty of this study lies in the fact that it is not only solving the Lane-Emden equation via PINNS but it is also determining the eigenvalue, $r$,  which is the stellar radius for each polytropic indices. By integrating the governing differential equation directly into the neural networks' loss function, PINNs provide a mesh-free, data-efficient, and flexible approach to determining the critical eigenvalues associated with self-gravitating polytropic spheres. Hyperparamter tuning were done using Bayesian optimization in Optuna framework. The results show that, for $n = 0, 1$, PINNs achieve near-exact agreement with theoretical eigenvalues (errors < $0.000806\%$). While for more nonlinear cases, $ n = 2, 3$ and $n=4$, PINNs yield errors below $0.0009\%$ and $0.05\%$ respectively, validating their robustness. 

Future studies could investigate hybrid methods combining PINNs with spectral techniques/heuristics algorithms to improve boundary condition handling. Exploring uncertainty quantification to assess the reliability of PINNs predictions in the absence of exact solutions. Despite their mesh-free advantage, PINNs require thousands of epochs to converge, which can be computationally expensive compared to traditional spectral or finite-difference methods for simple geometries. The accuracy of eigenvalues heavily depends on the choice of network depth, optimizer and activation function settings, making it sensitive to hyperparameters. In summary, PINNs offers a promising, scalable, and physics-compatible framework for eigenvalue problems in mathematical physics, with potential extensions to even more complex astrophysical and engineering challenges.


\bibliographystyle{unsrt}
\bibliography{reference1.bib}

\end{document}